\documentstyle[12pt,mpi]{article}
\newcommand{\F}{$ F_{2}(x,Q^2)\,$} 
\newcommand{\FL}{$ F_{L}(x,Q^2)\,$}

\newcommand{\Fz}   {{{F}_3}}
\begin{document}
%
%
\title{STANDARD MODEL PHYSICS AT HERA} 
\author{Vladimir Chekelian (Shekelyan) \\
{\em MPI f\"ur Physik (Munich) and ITEP (Moscow) } \\
(on behalf of the H1 and ZEUS Collaborations)
}
\maketitle
\baselineskip=14.5pt
    \vspace{-2.0cm}
\begin{abstract}
Both components of the Standard Model, 
QCD and the electroweak sector, are tested in the H1 and ZEUS experiments
at HERA.
The inclusive $e^- p$~ and $e^+ p$~single and double differential 
cross sections for neutral and charged current processes are measured 
in the range of four-momentum transfer squared, $Q^2$,
between $0.045$ and $30\,000$ GeV$^2$, and Bjorken $x$ between $10^{-6}$
and $0.65$. 
The neutral current double differential cross section, 
from which the proton structure
  function \F and the longitudinal structure function \FL are
  extracted, is measured at low $x$ with typically 1\% statistical and 2-3\%
  systematic uncertainties.  
In a next-to-leading order (NLO) DGLAP QCD analysis 
  using the H1 measurements and the $\mu p $ data of
  the BCDMS collaboration the strong coupling constant $\alpha_s$ 
  and the gluon momentum distribution are simultaneously determined.  A
  value of $\alpha _s (M_Z^2) =0.1150 \pm 0.0017 (exp)
  ^{+0.0009}_{-0.0005}~(model)$ is obtained, with an
  additional theoretical uncertainty of about $\pm 0.005$, mainly due
  to the uncertainty of the renormalisation scale.
$F_2^c$, the charm contribution to $F_2$, is measured and well described
by the boson gluon fusion production mechanism 
using the gluon distribution from the NLO QCD analysis.
The gluon distribution and the strong coupling constant $\alpha_s$ 
are also determined from jet production. 
For $Q^2>1\,000$~${\rm GeV}^2$, 
an asymmetry between $e^+p$ and $e^-p$ neutral current scattering 
is observed and the structure function $x\Fz$ is extracted.
A fit to the charged current data is used to determine a value
for the $W$ boson propagator mass. All data are found to be in good
agreement with Standard Model predictions.
\end{abstract}
\vspace{0.5cm}
\begin{center}
\it{ Talk given at 
     XV Rencontres de Physique de la Vall\'ee d'Aoste, 
     La Thuile, Italy, \\
     March 4-10, 2001 }
\end{center} 
\vspace{-23cm} 
\begin{large}
\begin{flushleft}
{\hfill    MPI--PhE/2001--13} \\
{\hfill    July 2001}
\end{flushleft}
\end{large} 
\baselineskip=17pt
\newpage
\section{Introduction}
An understanding of the structure of the proton in terms of
partons, i.e. quarks and gluons
which follow QCD dynamics, 
was established to a large extent
in lepton-proton, in particular electron-proton, 
deep inelastic scattering (DIS).
At the HERA $ep$ collider with 
an electron-proton centre-of-mass energy of $\sqrt{s}$=320~GeV,
effects of the electroweak sector can also be tested in DIS
in the region where 
the squared four momentum transfer
$Q^2 \simeq M_Z^2$ or $M_W^2$
($M_Z$ and $M_W$ are the $Z^{\rm o}$ and $W$ boson masses).
In addition, signals of new physics beyond the Standard Model 
may be expected to arise at the highest $Q^2$ where 
the proton structure is probed at smallest distances.

\begin{figure}[h]
    \vspace{1.0cm}
  \begin{minipage}[t]{0.40\textwidth}
      \makebox[0cm][t]{}        
      \begin{center}
\includegraphics{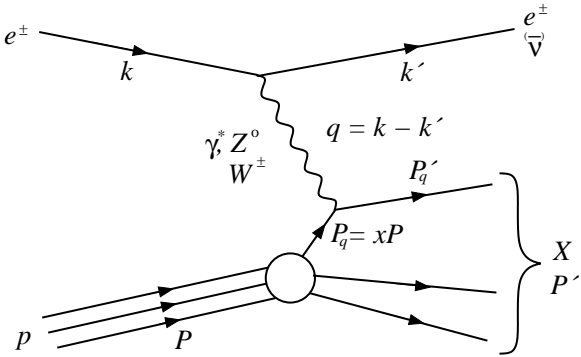}
      \end{center}
  \end{minipage}
  \hfill
  \begin{minipage}[t]{0.45\textwidth}
    \vspace{-0.5cm}
    \caption[]{\it Feynman diagram for deep inelastic scattering (DIS).
The kinematics is defined by the four-vectors of the incoming proton, $P$,
the incoming electron, $k$, the scattered electron, $k^\prime$, 
and the exchanged boson, $q = k-k^\prime$.
}  
    \label{fig1}
    \vspace{1.0cm}
  \end{minipage}
\end{figure}

Deep inelastic scattering, Fig.~\ref{fig1}, depends on
three variables, $Q^2$, $x$ and $y$, where $Q^2=-q^2$ and
$x=Q^2/2(Pq)$ is the fraction of
the proton momentum carried by the struck quark 
and $y=(Pq)/(Pk)$ is the inelasticity of the interaction. 
The variables are related via $Q^2=sxy$.

The electron-proton neutral current (NC) and charged current (CC) 
cross sections can each be expressed in terms of
three proton structure functions.
For neutral current interactions
mediated by a $\gamma$ or $Z^{\rm o}$~boson,
the proton structure function $ F_{2}(x,Q^2)\,$
dominates the cross section and
is proportional to the sum of the quark momentum distributions weighted
by the quark charge squared.
The longitudinal structure function $ F_{L}(x,Q^2)\,$ is
directly sensitive to the gluon momentum distribution. 
The structure function $ xF_{3}(x,Q^2)\,$  is related to valence quarks only. 
These three proton structure functions 
have been measured at HERA and confronted with
the Standard Model predictions. The gluon momentum distribution and the strong
coupling $\alpha_s$ determined in a QCD analysis
of the HERA inclusive data have been cross checked 
in more exclusive channels such as charm and
jet production via boson gluon fusion. 
The propagator mass $M_W$ has been determined from the $Q^2$
dependence of the CC interactions and can be compared with
direct $M_W$ mass measurements in $e^+e^-$ and $p \bar p$ interactions. 
Thus, measurements at HERA provide tests of both components
of the theory, the electroweak sector 
and the evolution of the parton densities
as predicted by perturbative QCD.

\begin{figure}[h]
    \vspace{8.0cm}
  \begin{minipage}[t]{0.40\textwidth}
      \makebox[0cm][t]{}        
      \begin{center}
\includegraphics{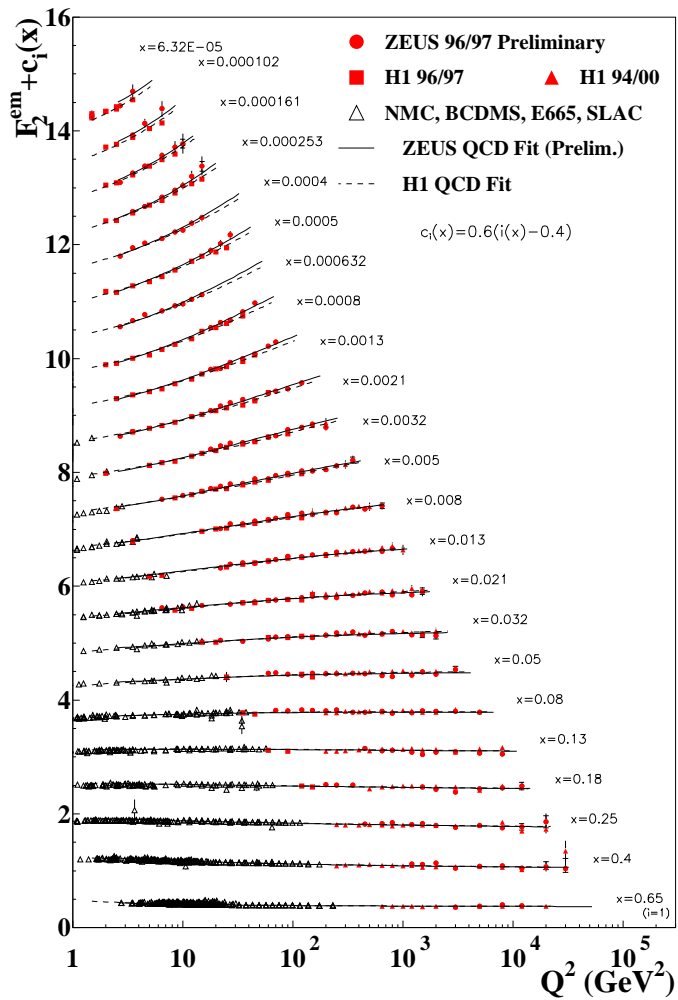}
      \end{center}
  \end{minipage}
  \hfill
  \begin{minipage}[t]{0.35\textwidth}
    \vspace{-2.0cm}
    \caption[]{\it 
Measurement of the proton structure function $F_{2}(x,Q^2)$
by H1 and ZEUS and by fixed target 
charged lepton-proton experiments. The H1 and ZEUS NLO QCD fits 
are also shown.}
    \label{f2}
    \vspace{3.5cm}
  \end{minipage}
\end{figure}

\begin{figure}[h]
    \vspace{10.0cm}
  \begin{minipage}[h]{0.40\textwidth}
      \makebox[0cm][h]{}        
      \begin{center}
\includegraphics{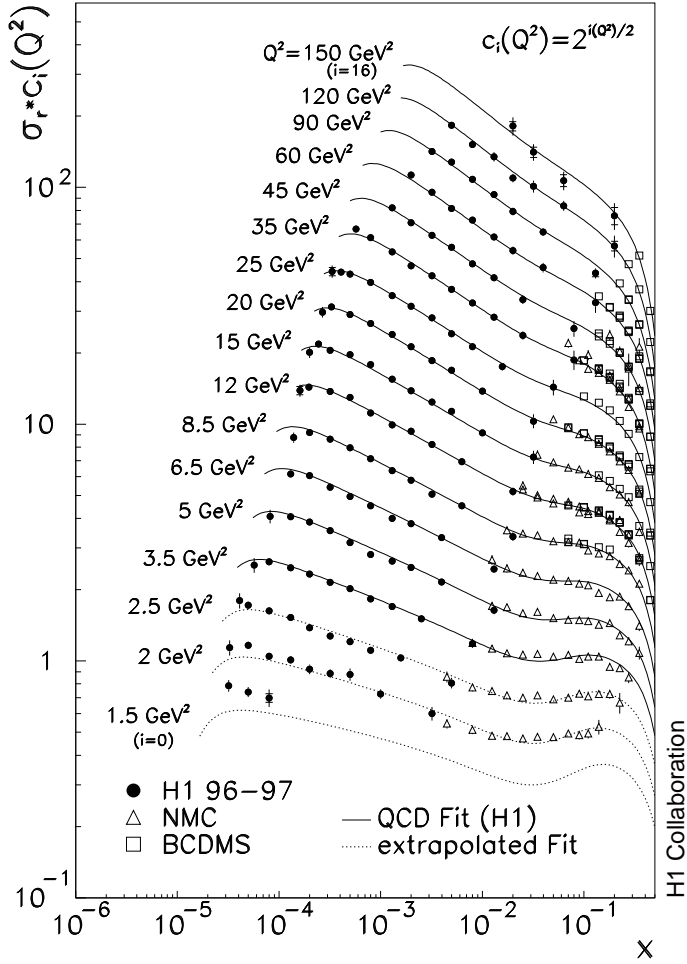}
      \end{center}
  \end{minipage}
  \hfill
  \begin{minipage}[h]{0.35\textwidth}
    \vspace{-6.0cm}
    \caption[] {\it 
   Measurement of the NC reduced cross section.
   The solid curves show the result of 
   a QCD fit to the H1 data with $Q^2 \geq 3.5$GeV$^2$.
   The dashed curves  show the extrapolation of this fit towards lower $Q^2$.}
    \label{redsec}
  \end{minipage}
    \vspace{2.0cm}
\end{figure}

\section{Results}
The first $F_2$ measurements\cite{h1f292} 
in $ep$ interactions at HERA, based on a luminosity of 
0.025 pb$^{-1}$ collected in 1992, revealed at low $x$
a pronounced rise of $F_2$ with decreasing $x$
which generated a new interest in QCD in DIS.
Since then the collected luminosity has been increased 
by more than three orders
of magnitude. In addition, the H1 and ZEUS detectors have
been continuously improved.
In 1995/1996 new detector components were installed by H1:
the backward calorimeter Spacal, 
the backward drift chamber BDC and the backward silicon tracker BST.
In 1995 the ZEUS collaboration installed a beam pipe calorimeter (BPC)
and in 1997 a beam pipe tracker (BPT) was added, 
covering very small angles of the scattered electron.
These improvements and the increase in luminosity made it possible 
to extend the measured region\cite{h1f294,zeusf294,h1hiq2,zeushiq2} 
to very low $Q^2\geq0.045$~GeV$^2$,
covering the transition region from DIS to real photoproduction ($Q^2=0$),
as well as  very high $Q^2\geq10^4$~GeV$^2$.
The variable $x$ ranges from $10^{-6}$ 
to 0.65 in the valence quark region.
$ F_{2}$ measurements\cite{pub97,zeusf297} 
at HERA together with fixed
target experiments are shown in Fig.~\ref{f2} as a function
of $Q^2$ for different values of $x$. The H1 and ZEUS data agree
very well in the full range accessible by the HERA experiments.  
At very low $y=0.003$, the HERA data overlap and agree with the data of 
the fixed target experiments.
In a large part of the phase space 
the precision of the data is $2-3\%$ which is comparable with
that of fixed target experiments. 

%
\begin{figure}[t] 
\vspace{12.1cm}
\includegraphics{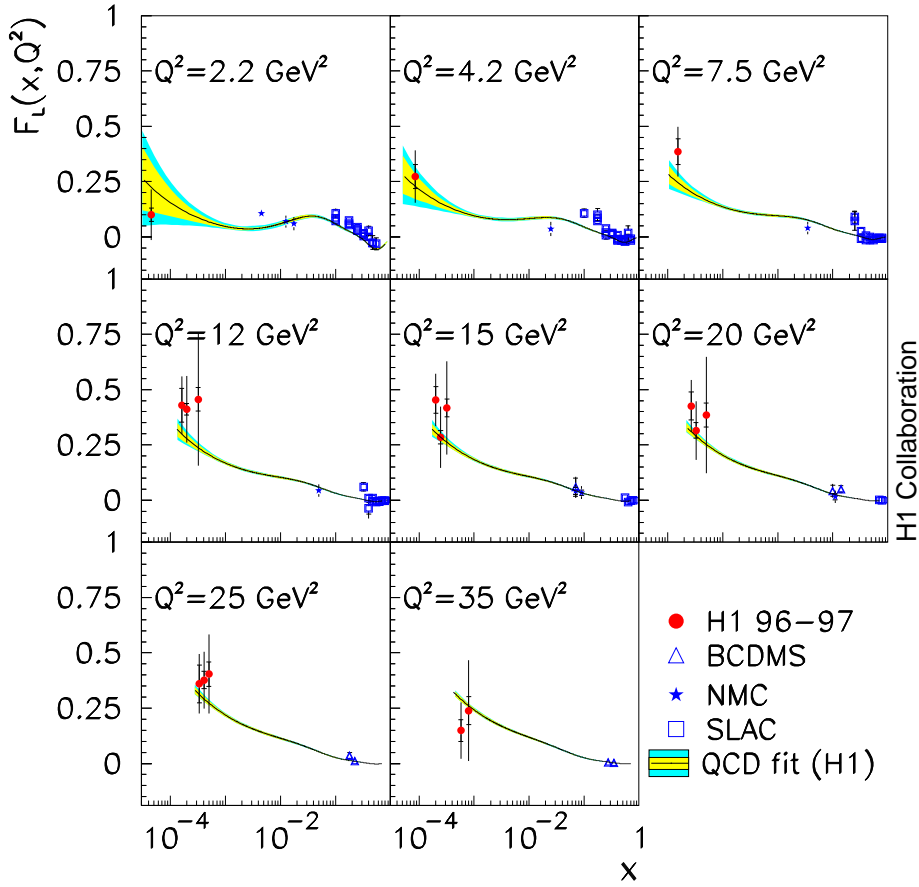}
\caption {\it 
    The longitudinal structure function $ F_{L}(x,Q^2)\,$
    for different bins of $Q^2$  as obtained
    by H1 at low $x$, and by charged lepton-nucleon fixed
    target experiments at large $x$.
    The outer (inner) error
    bars show the total (statistical) errors. 
    The error bands are due to the
    experimental (inner) and model (outer) uncertainty of the
    calculation of $F_L$ using the QCD fit to the H1 
    data for $y < 0.35$ and $Q^2 \geq  3.5$~GeV$^2$. 
\label {fl} }
\end{figure}
\begin{figure}[t] 
\vspace{10.0cm}
\includegraphics{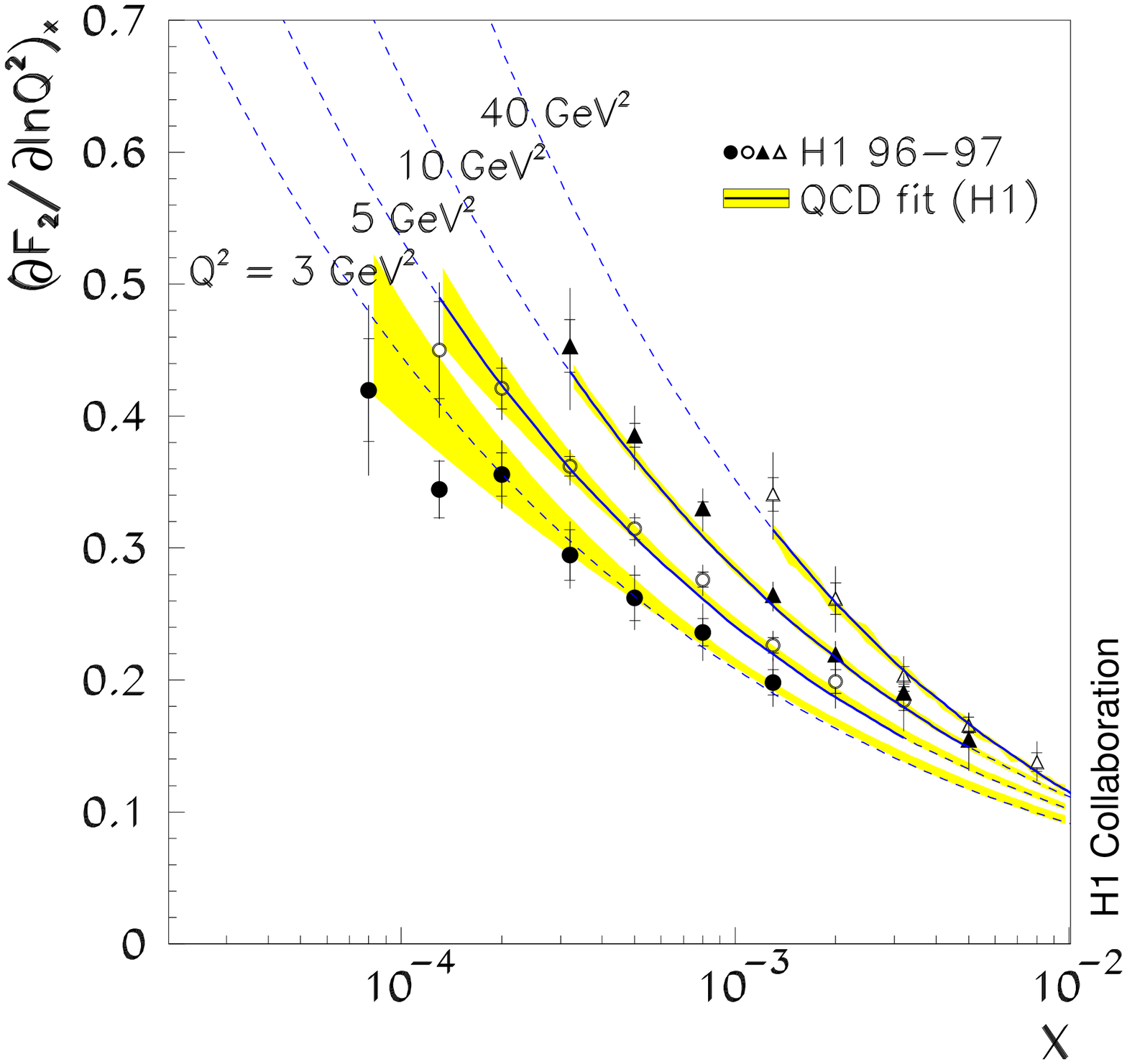}
\includegraphics{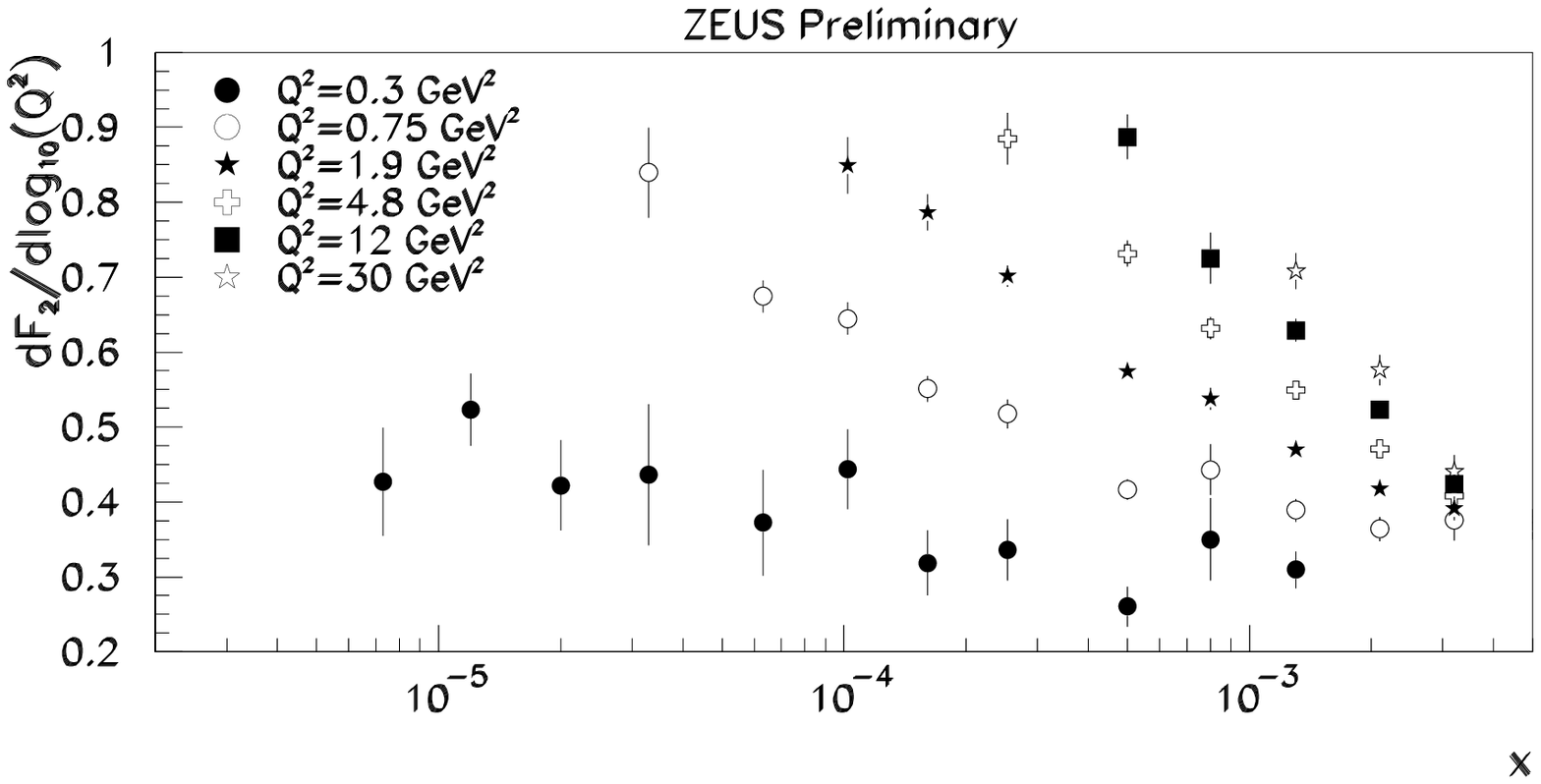}
\includegraphics{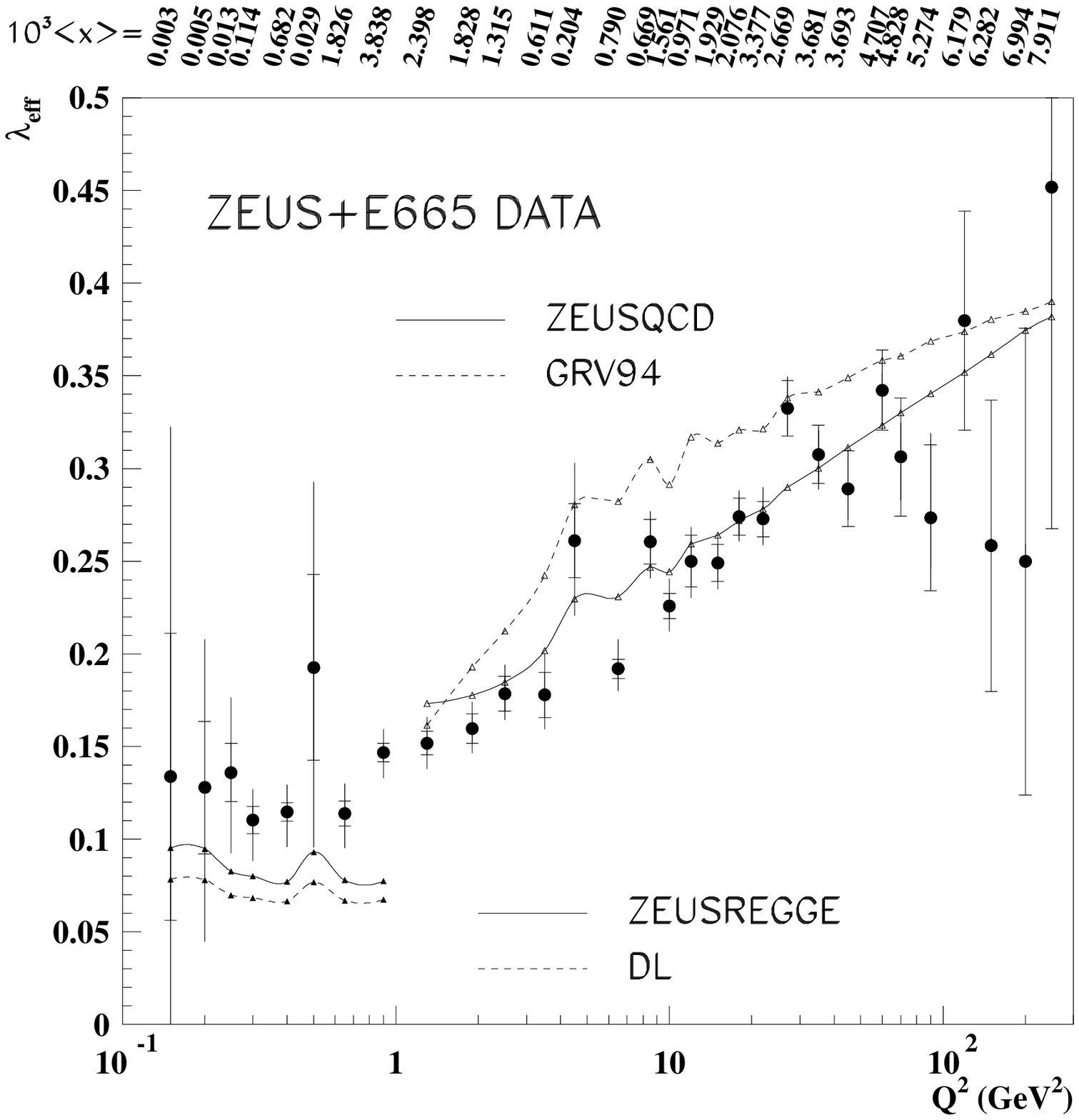}
\begin{picture}(0,0)
\put(185,200){(a)}
\put(185,75){(b)}
\put(250,180){(c)}
\end{picture}
\caption {\it 
 (a),(b) The derivative $(\partial F_{2} / \partial \ln  Q^{2})_x\,$  plotted
 as functions of $x$ for different $Q^2$,
 for the H1 and ZEUS data  (symbols).
 The QCD fit to
 the H1 data for $Q^2 \geq 3.5$GeV$^2$ is shown by solid lines. 
 The dashed curves extrapolate this fit.
 The error bands represent the model uncertainty of the QCD analysis. 
(c) $\lambda_{eff}=d\ln F_2/d\ln(1/x)$ as a function of $Q^2$ calculated
by fitting $F_2=A x^{-\lambda_{eff}}$ to ZEUS and E665 data with $x<0.01$. 
The inner error bar shows the statistical error and the outer the total 
statistical and systematic error added in quadrature.  
\label {deriv}    }
\end{figure}
\begin{figure}[h] 
\vspace{12.5cm}
\includegraphics{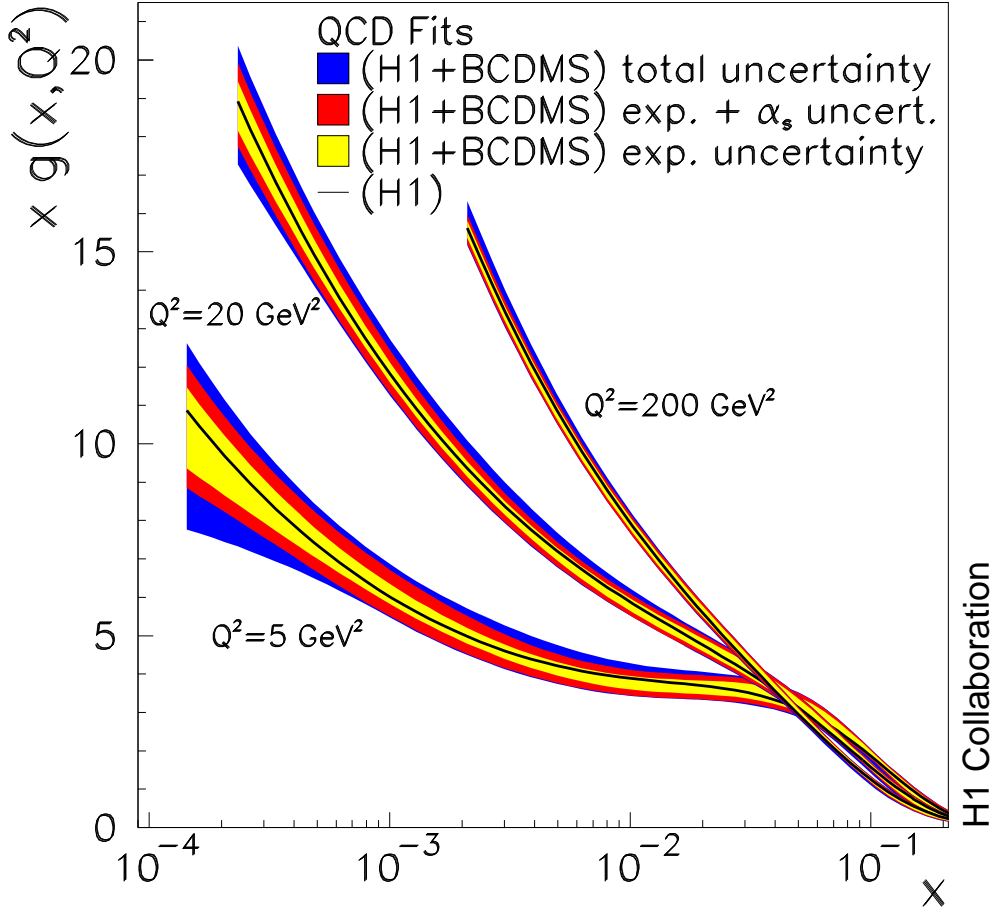}
\caption {\it 
   Gluon distribution resulting from the QCD fit 
   to H1  $ep$ and  BCDMS $\mu p$ data.
   The innermost error bands represent the experimental
   error for fixed $\alpha_s(M_Z^2)\,$=0.1150. 
   The middle error bands include in addition
   the contribution due to the simultaneous fit of $\alpha_s\,$.
   The outer error bands also include  the model  
   uncertainties of the QCD analysis and the dependence
   on the data range fitted.
\label {xg} }
\end{figure}

\begin{figure}[p] 
\vspace{9cm}
\includegraphics{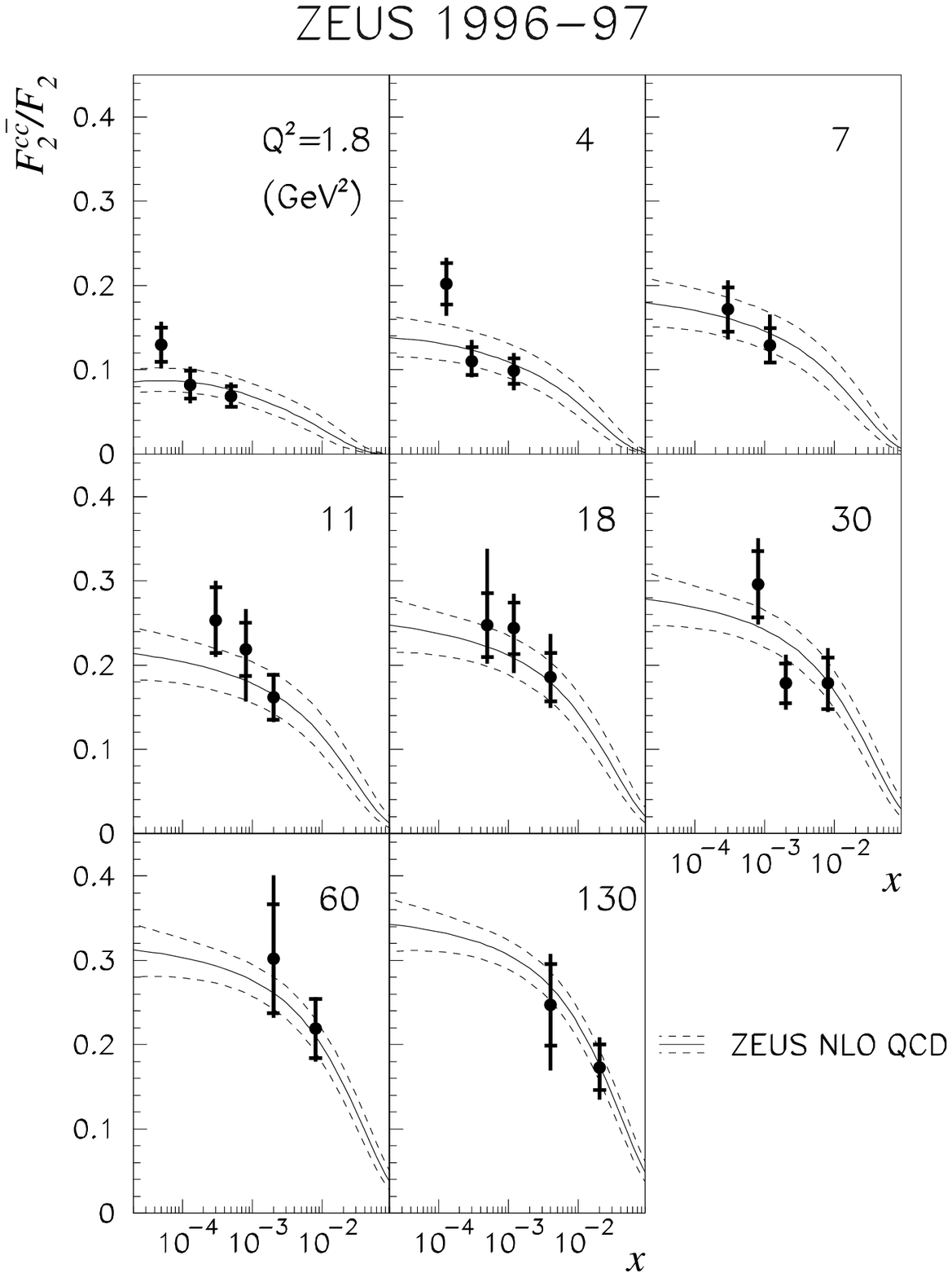}

\includegraphics{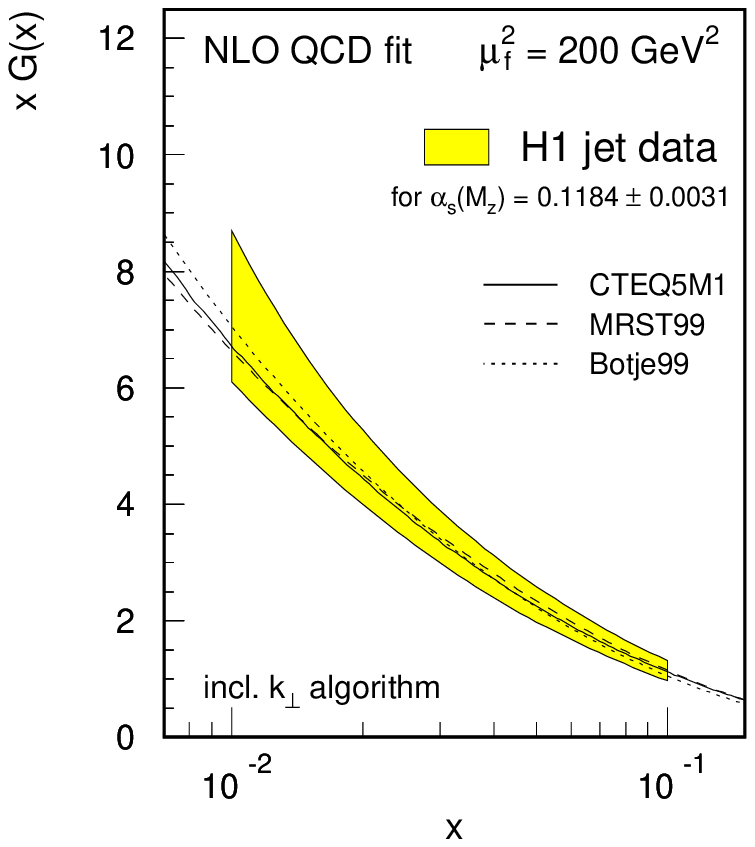}
\begin{picture}(0,0)
\put(165,220){(a)}
\put(400,100){(b)}
\end{picture}
\caption {\it 
Charm contribution to $F_2$ in comparison with
calculations using the NLO QCD fit to $F_2$ (a).
Gluon distribution resulting from jet studies (b).
The error band includes the experimental and the theoretical
uncertainties as well as the uncertainty of
$\alpha_s(M_Z^2)$.
\label {f2c} }
%
\vspace{8.0cm}
\includegraphics{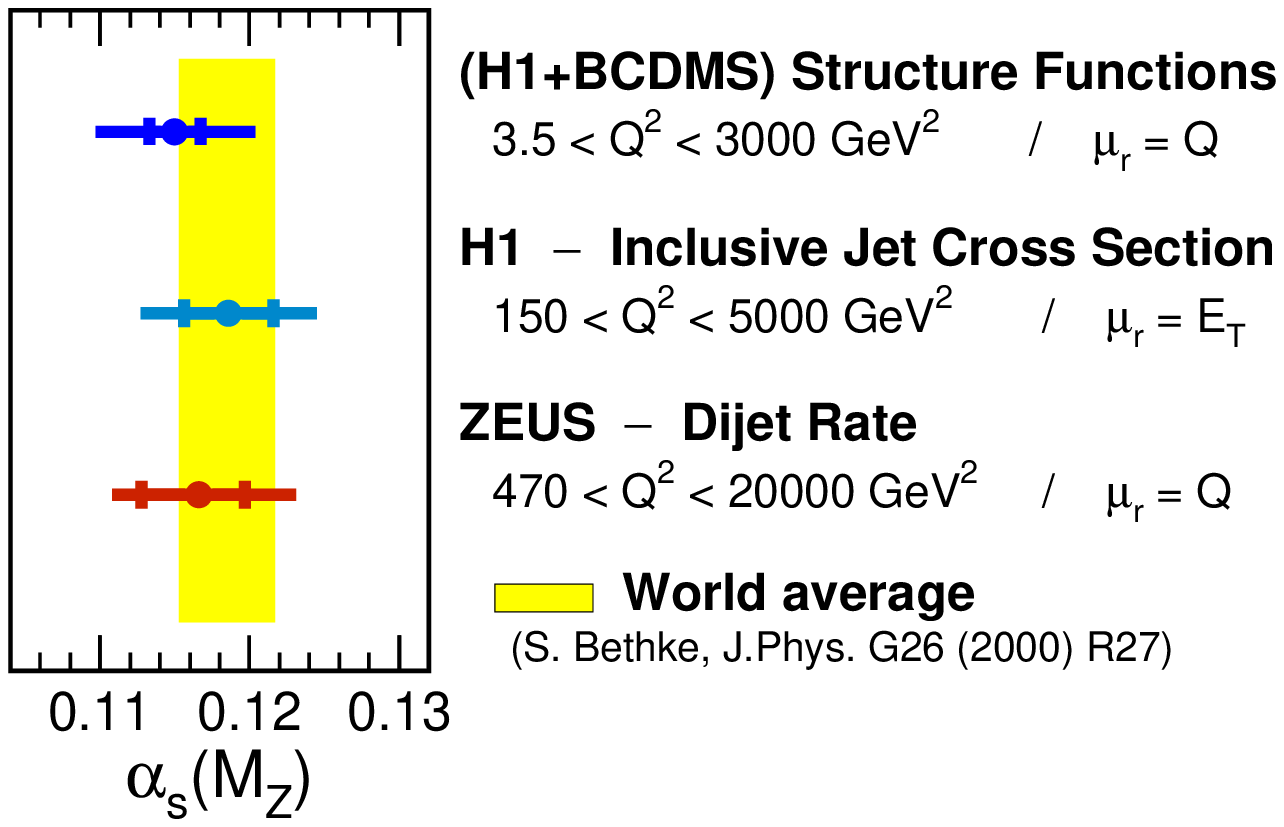}
\caption {\it 
Measurements of the strong coupling constant $\alpha_s(M_Z^2)$ 
at HERA in comparison with the world average.
Inner error bars represent
experimental errors, outer error bars represent
total errors including the theoretical uncertainty
mainly due to
the uncertainty of the renormalisation and the factorisation scales.
\label {alphas} }
\end{figure}

The results on the reduced NC cross section, defined as the cross section
divided by the kinematical factor $(2 \pi \alpha^2 Y_+)/(Q^4 x)$
with $Y_+ = 1+ (1-y)^2$ and $\alpha$, the fine structure constant, 
are shown in Fig.~\ref{redsec} as function of $x$ 
for $1.5 \leq Q^2 \leq 150$~GeV$^2$.
At very high $y$ ($y\simeq 0.82$), i.e. the lowest x points at fixed $Q^2$, 
a turn over is visible in the measured cross section
which is due to the longitudinal structure function $F_L(x,Q^2)$.
From these high-y cross sections,
$F_L(x,Q^2)$ was extracted\cite{pub97} as shown in Fig.~\ref{fl}.
This extends the knowledge of the longitudinal structure
function to much lower $x$ than available
from fixed target lepton-proton scattering experiments.
The increase of $F_{L}(x,Q^2)\,$ 
towards low $x$ is consistent
with the QCD calculation, reflecting the rise
of the gluon momentum distribution towards low~$x$. 

Another quantity related to the gluon density is 
the derivative $(\partial F_{2} / \partial \ln  Q^{2})_x\,$
which is shown 
as a function of $x$ in Fig.~\ref{deriv}a,~b for different $Q^2$.  
The derivatives show a continuous rise towards low
$x$ for fixed $Q^2$ which is well described by the QCD fit to
the H1 data down to $Q^2 = 3$ GeV$^2$.
The shape of $(\partial F_{2} / \partial \ln  Q^{2})_x\,$
reflects the behaviour of the gluon distribution in the associated 
kinematic range.

The rise of $F_2$ with decreasing  $x$ can be characterised
by $F_2 \propto x^{-\lambda_{eff}}$ 
where $\lambda_{eff}=d\ln F_2/d\ln(1/x)$.
In Fig.~\ref{deriv}c, values of 
$\lambda_{eff} $ are shown from fits 
at fixed $Q^2$ to the $F_2$ data. 
The prominent rise of $F_2$ at large $Q^2$ values
is in agreement with the QCD fit\cite{zeuslowq2}.  
For smaller $Q^2$, the rise of $F_2$ becomes less steep and
$\lambda_{eff} $ approaches a value of ~$\approx0.08$ expected from the
Regge motivated fits\cite{zeuslowq2,dola}.

The high precision of the data allows
a simultaneous determination of the gluon distribution and 
the strong coupling constant $\alpha_s\,$
in a next-to-leading-order (NLO) DGLAP\cite{DGLAP} QCD fit\cite{pub97}
by combining the low\,$x$ cross section data of H1 
with $\mu p$ scattering data of the BCDMS collaboration\cite{BCDMS}
at large $x$.
The gluon distribution from this fit is shown in Fig.~\ref{xg} 
for $Q^2$=5, 20, and 200 GeV$^2$.
The inner error band represents the experimental uncertainty of the
determination of $xg$ for $\alpha_s\,$ fixed. 
The simultaneous determination of $xg(x,Q^2)\,$ and $\alpha_s\,$ 
leads to a small increase of the
experimental error of $xg\,$ as is illustrated by the middle error band.
The full error band includes in addition the uncertainties connected
with the fit ansatz (``model uncertainty'').
The DGLAP analysis of the data
leads to a gluon distribution which rises dramatically at
small $x$ with increasing $Q^2$. 
A value of the strong coupling constant 
$\alpha_s(M_Z^2)\,$ = $0.1150 \pm 0.0017
(exp)^{+~~0.0009}_{-~~0.0005}(model)$ is obtained. 
The value of $\alpha_s\,$
changes by about 0.005, much more than the experimental error, if the
renormalisation scale is allowed to vary by a factor of four in the fit.
This theoretical uncertainty is expected to be reduced significantly in
next-to-next-to-leading-order (NNLO) perturbation theory.  

\begin{figure}[t] 
\vspace{8.7cm}
\includegraphics{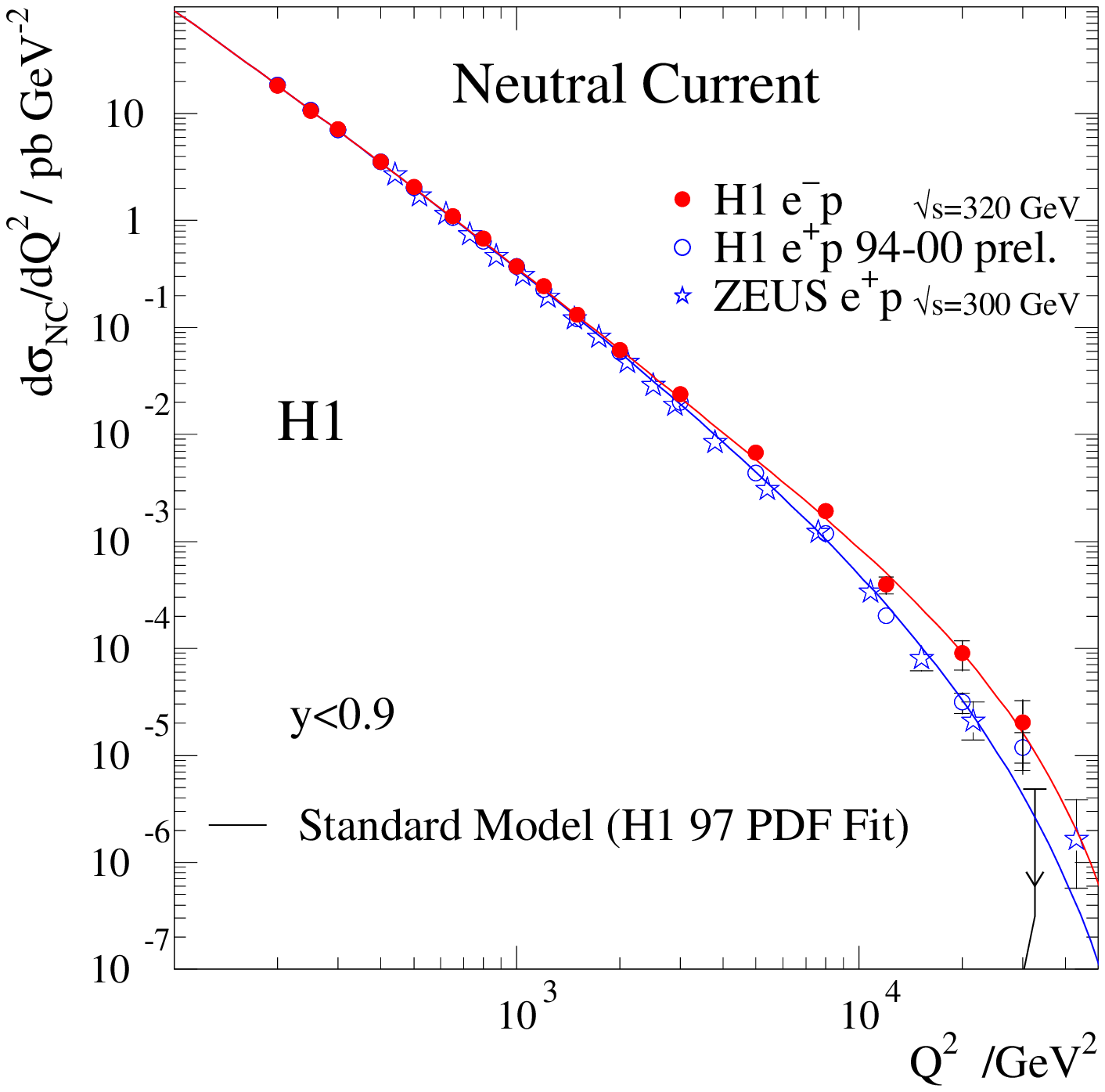}
\includegraphics{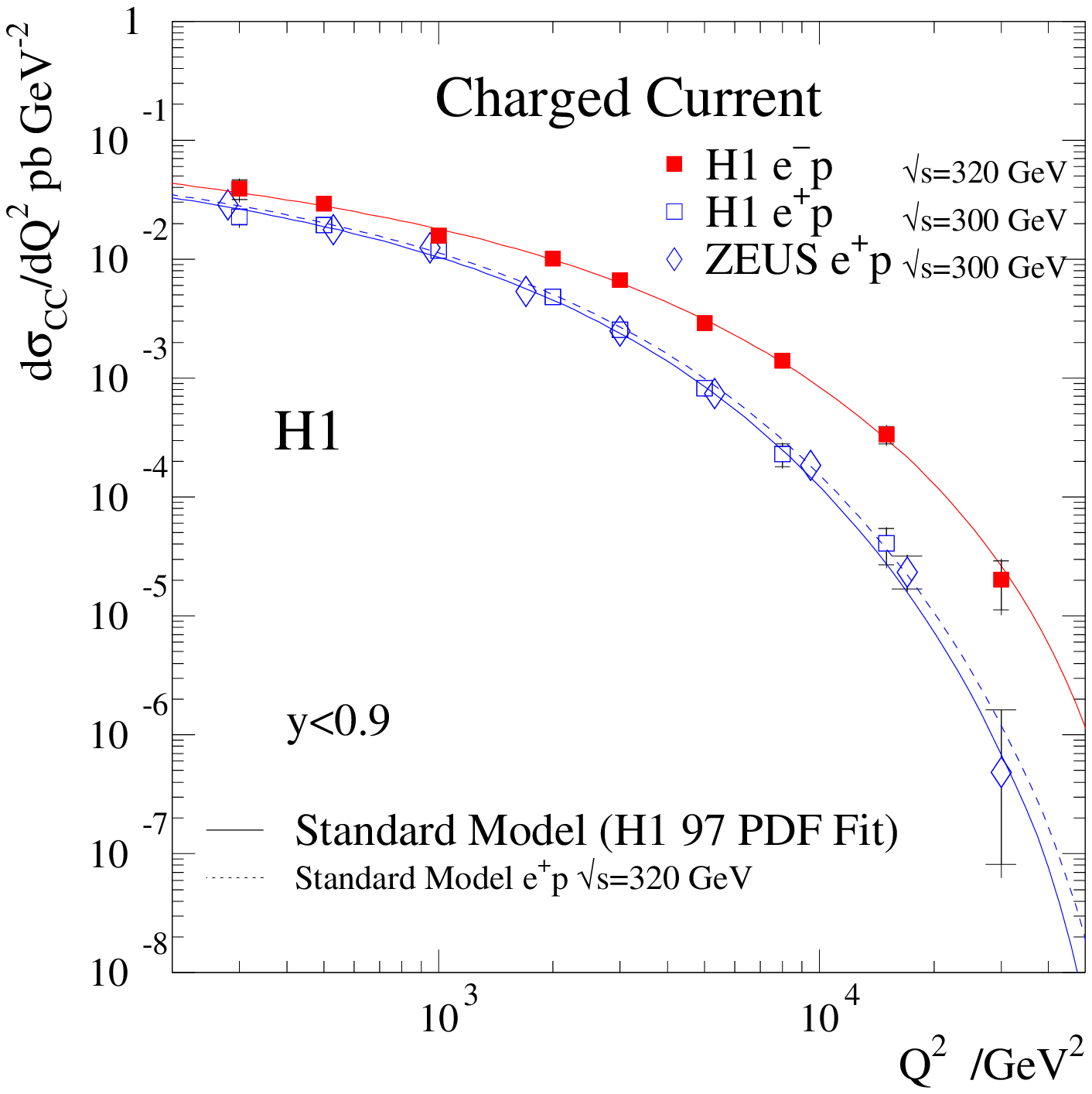}
\begin{picture}(0,0)
\put(185,200){(a)}
\put(408,200){(b)}
\end{picture}
\caption {\it 
   The $Q^2$ dependences of the NC (a) and CC (b) cross sections 
   are shown for the $e^-p$ (solid symbols) and
   $e^+p$ data (open symbols).
   The data are compared with the Standard Model
   expectations determined from the NLO QCD fit. 
\label {dsdq2} }
\end{figure}

Exclusive channels such as charm or jet
production via boson gluon fusion
can also be used for a direct extraction of the gluon density.
The charm contribution to $F_2$, $F_2^c$, is determined at HERA using
measurements of $D^*$ production and account for
$25-30\%$ of $F_2$\cite{f2c} as shown in ~Fig.~\ref{f2c}a. 
The $F_2^c$ results are in agreement with
expectations calculated using the gluon momentum distribution
as determined in the NLO QCD fit to the inclusive cross sections.
Jet production data\cite{jets} could be used to determine
the gluon distribution using $\alpha_s$ from other experiments,
or determine the strong coupling constant $\alpha_s$ using parton momentum
distributions from global QCD fits.
The resulting gluon distribution at the
factorisation scale $\mu_f = \sqrt {200}$~GeV 
is shown in Fig.~\ref{f2c}b 
and consistent with the gluon from the global QCD fits.
The results for $\alpha_s$ from jet studies
by H1 and ZEUS collaborations
together with $\alpha_s$ determined from scaling violations by H1
are shown in Fig.~\ref{alphas} in comparison with
the world average\cite{bethke}. 

\begin{figure}[t]
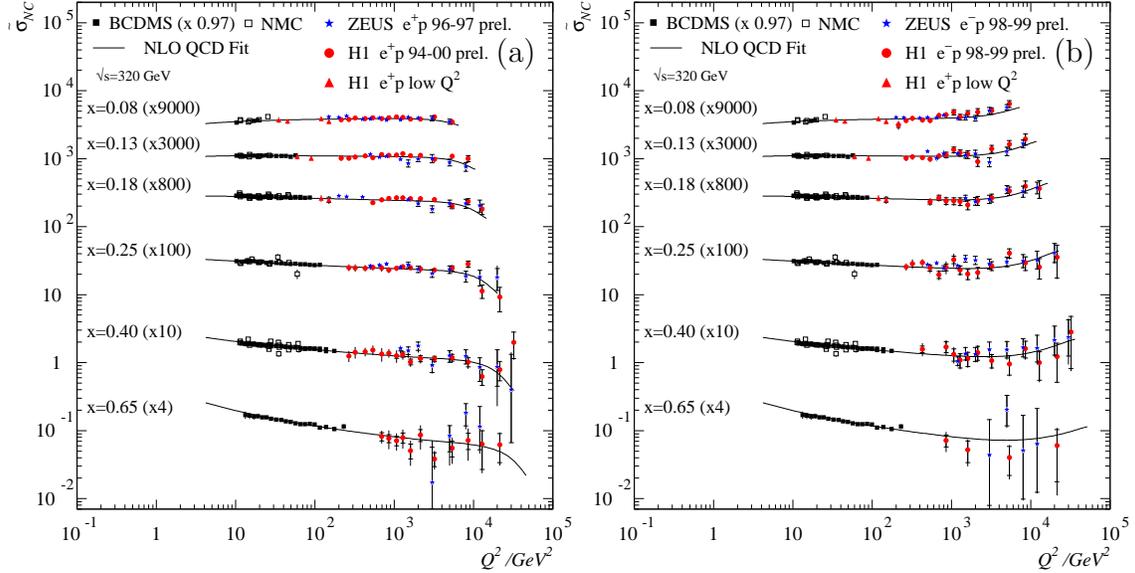

\vspace{9.0cm}
\includegraphics{fig/redxsec_e+_zeus_h1comb.epsi}
\includegraphics{fig/redxsec_e-_zeus_h1.epsi}
\begin{picture}(0,0)
\put(193,205){(a)}
\put(404,205){(b)}
\end{picture}
\caption {\it 
  The NC $e^+p$ (a) and $e^-p$ (b) reduced cross sections 
  are shown at high $x$ 
  together with fixed target data from
  BCDMS and NMC.  The curves represent the
  Standard Model expectation based on the NLO QCD Fit.
\label {redsechix} } 
\end{figure}

\begin{figure}[p]
\vspace{8.0cm}
\includegraphics{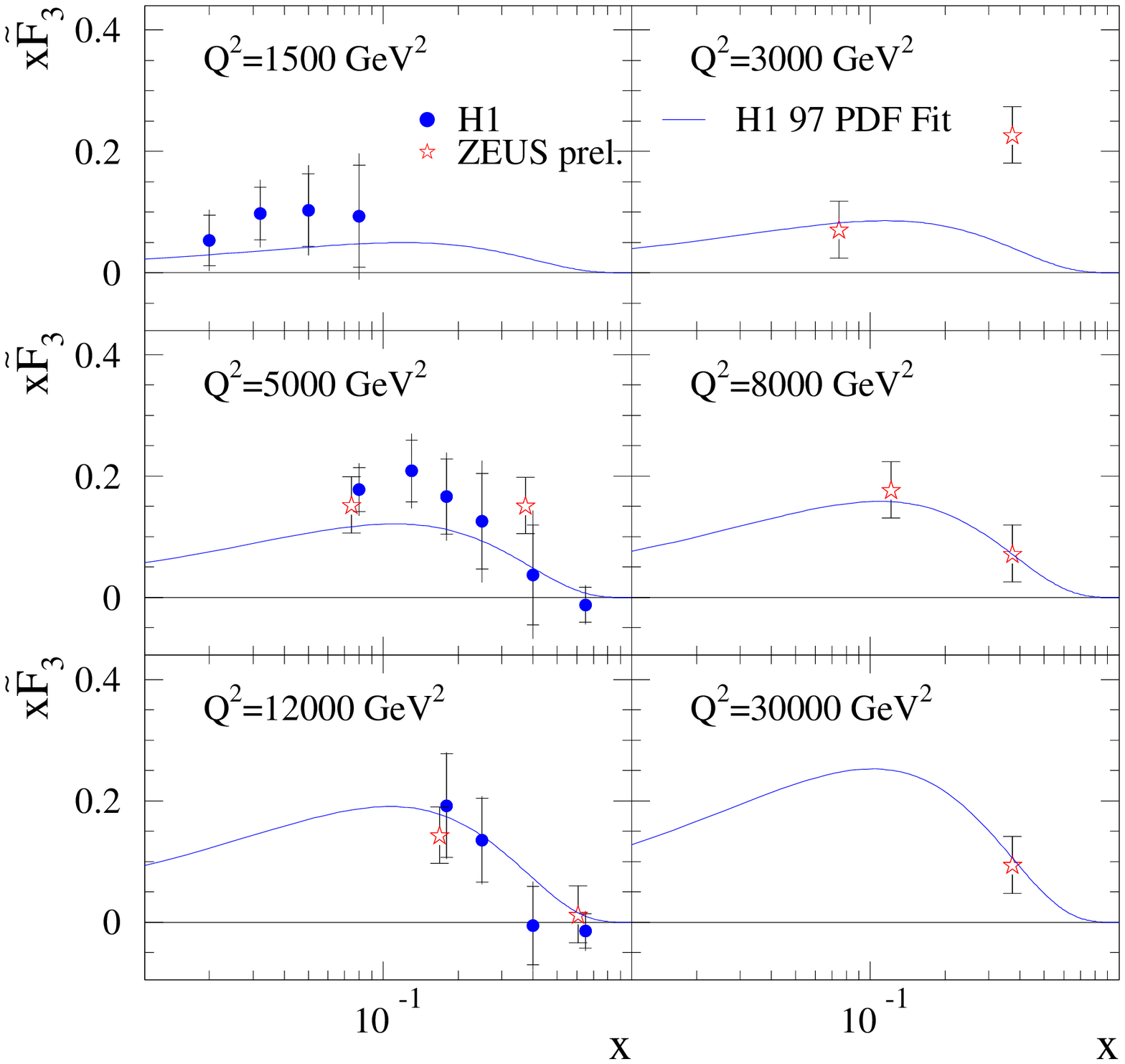}
\includegraphics{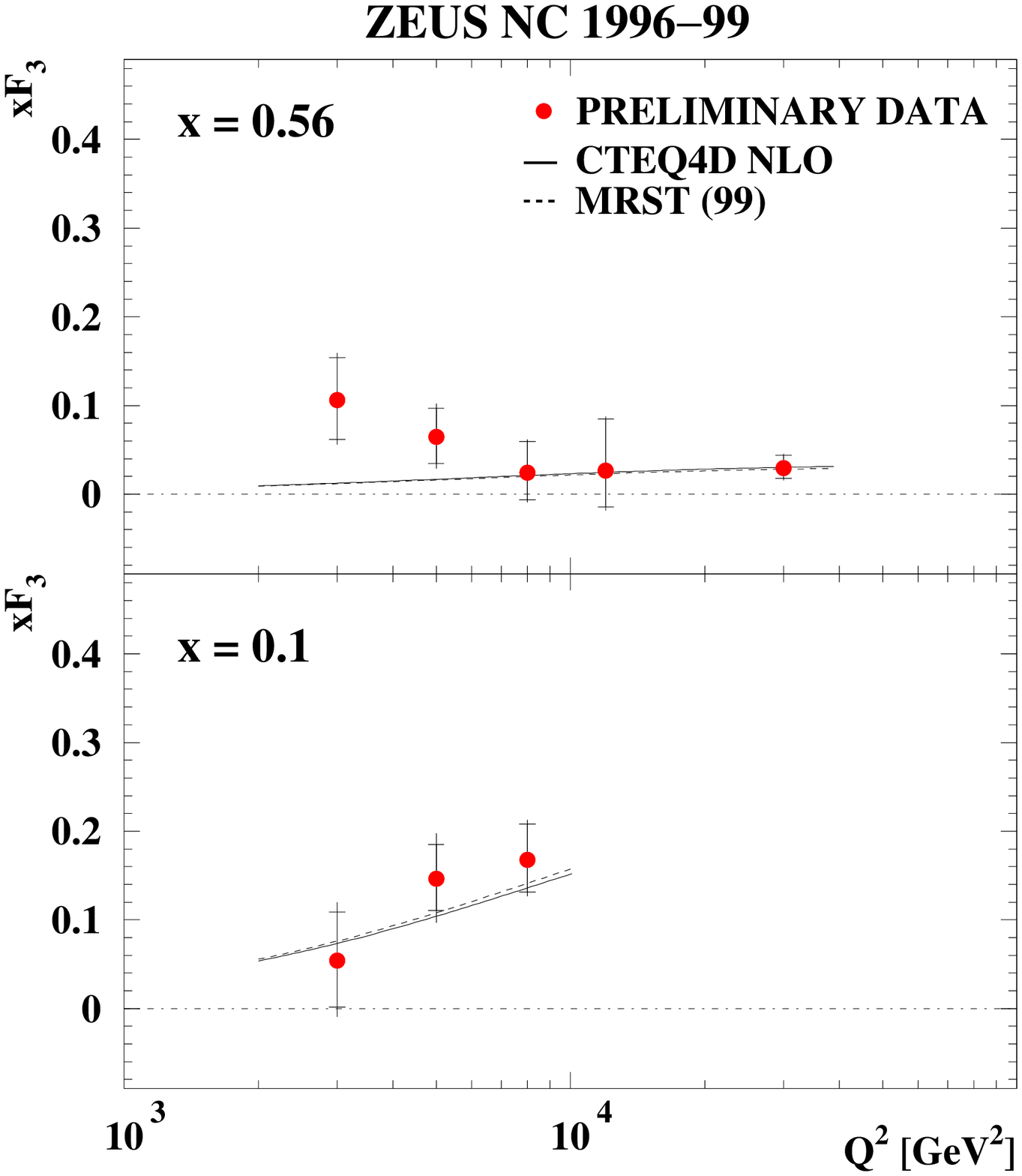}
\begin{picture}(0,0)
\put(35,200){(a)}
\put(270,200){(b)}
\end{picture}
\caption {\it 
The structure function $x\Fz$ measured by H1 and ZEUS is compared 
with the H1 97 PDF Fit\cite{h1hiq2} and global fits\cite{cteq,mrs}.
\label {xf3} }
%
\vspace{8.0cm}
\includegraphics{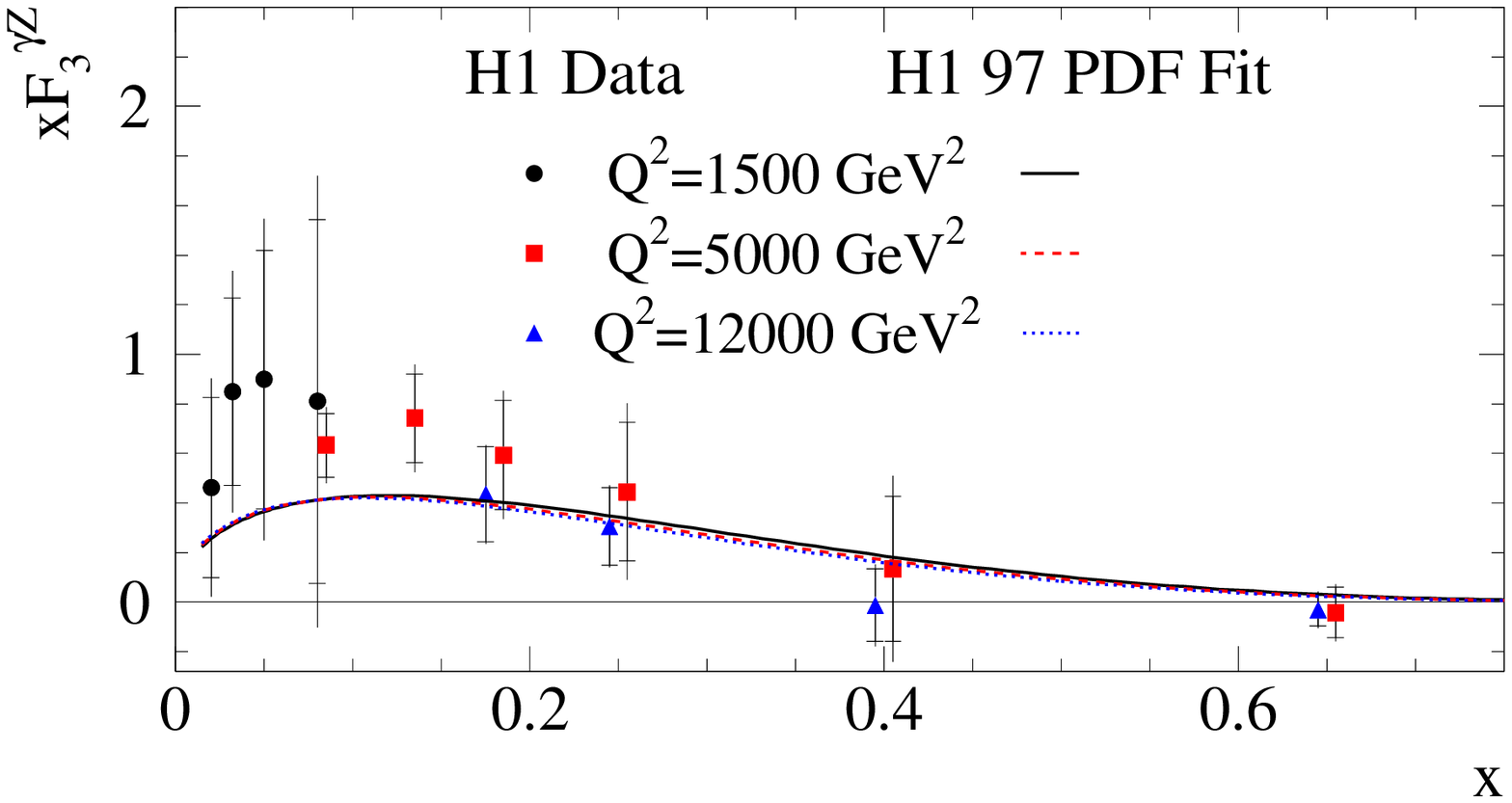}
\caption {\it 
   The structure function $x{F_3^{\gamma Z}}$ measured 
   by the H1 collaboration is compared with the H1 97 PDF Fit\cite{h1hiq2}.
\label {xf3gz} }
\end{figure}

\begin{figure}[h]
\vspace{9.5cm}
\includegraphics{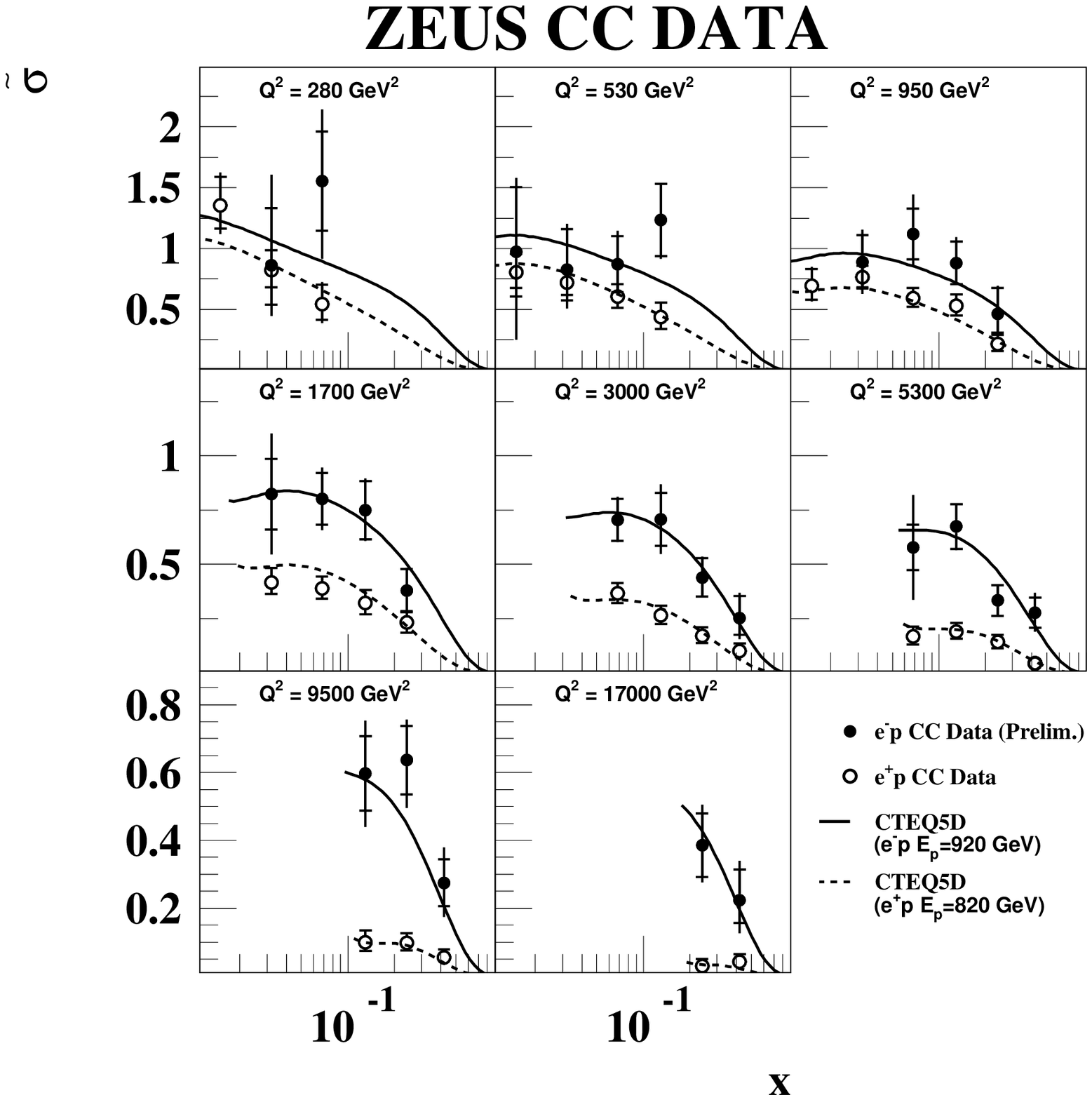}
\includegraphics{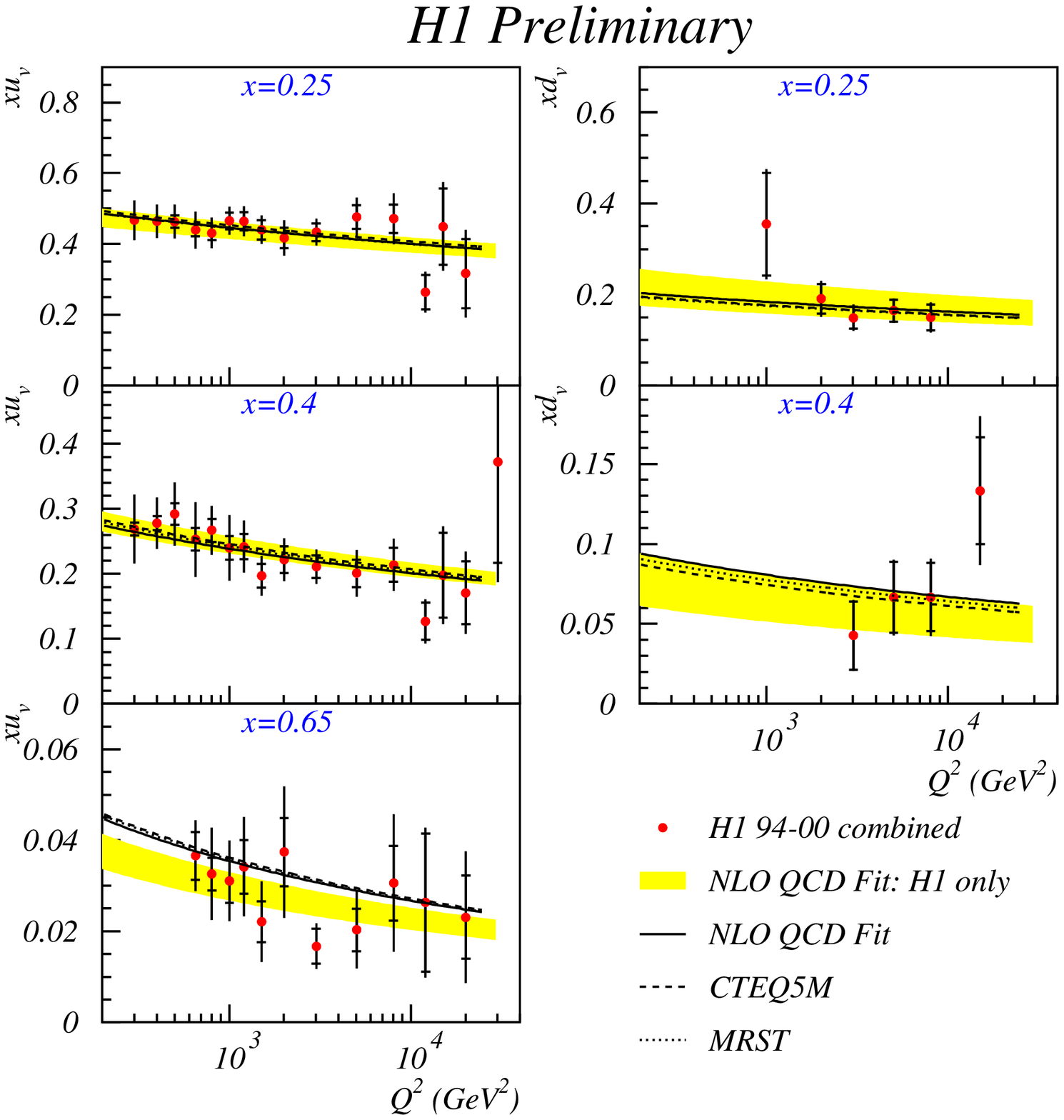}
\begin{picture}(0,0)
\put(185,210){(a)}
\put(408,210){(b)}
\end{picture}
\caption {\it 
The CC $e^{\pm}$ reduced cross sections are compared
with global QCD fit (a).
The valence quarks distributions $xu_v$ and $xd_v$
determined using a local extraction method
in comparison with the H1 NLO QCD and global fits\cite{cteq,mrs}(b).
\label {ccxsec} }
\end{figure}

Purely electroweak effects become observable at HERA
in a region of very high 
$Q^2$ comparable with the masses squared of the vector bosons
$Z^{\rm o}$ and $W$.
Especially interesting in this respect is the comparison of 
neutral and charged current cross 
sections\cite{h1hiq2,zeushiq2,epub,zeusnc98,l00p}
shown in Fig.~\ref{dsdq2}.
At low $Q^2$, the cross section of
the CC process, which is due to exchange of the $W$ boson,
is suppressed compared to that of the NC process 
due to the different propagator terms.
At high  $Q^2\approx M_W^2$, 
both cross sections are approaching each other
manifesting the unification of the weak and electromagnetic forces. 

Electroweak effects are also visible in the comparison of 
the $e^-p$ and $e^+p$ NC cross sections.
Due to $Z^{\rm o}$ exchange,  
at very high  $Q^2$ the NC cross section is predicted to be suppressed
in $e^+p$ and enhanced in $e^-p$ interactions.
This effect becomes visible for $Q^2 \geq $1000~GeV$^2$
as shown in Fig.~\ref{dsdq2}a,~\ref{redsechix} and 
agrees with the Standard Model prediction.

The difference of the NC $e^+p$ and $e^-p$ cross sections at large $Q^2$
has been used to extract the structure function $x{{F}_3}$~
in the range $0.02<x<0.65$ and $1\,500 \leq Q^2 \leq 30\,000$ GeV$^2$
as shown in ~Fig.\ref{xf3}.
Since at HERA the dominant contribution to $x{{F}_3}$ is due to
photon-$Z^{\rm o}$ interference, the H1 collaboration
evaluated the corresponding structure function $x{F_3^{\gamma Z}}$.
The result is shown in Fig.~\ref{xf3gz}. 
The integral $\int_{0.02}^{0.65} {F_3^{\gamma Z}} {\rm d}x = 
1.88 \pm 0.35 (\rm stat.) \pm 0.27 (\rm syst.)$ 
is found to be consistent within experimental errors
with theoretical expectation of 1.11\cite{epub}.

The reduced CC cross sections, defined in analogy to the NC case 
as the cross section divided by 
$(G_F^2 M_W^4)/(2 \pi x (M_W^2+Q^2)^2)$ with
the Fermi coupling constant $G_F$, are shown
in Fig.~\ref{ccxsec}a. The results are well described by the 
Standard Model expectations. 
At high $x$ the $e^-p$ and $e^+p$ cross sections
are dominated by the valence quark contributions,
$xu_v$ and $(1-y)^2 xd_v$ respectively.
For points where the expected $xq_v$ contribution to the total cross section
is larger than 70\% a local extraction method\cite{l00p} is used 
to determine the valence quark densities:
$$ xq_v(x,Q^2)=
\sigma_{meas}(x,Q^2)
\left(\frac{xq_v(x,Q^2)}{\sigma(x,Q^2)}\right)_{theory}$$
where $\sigma_{meas}$ is the measured NC or CC $e^{\pm}p$ double differential
cross sections and the ratio is the theoretical expectation
from the H1 97 PDF Fit\cite{h1hiq2}. The extracted parton densities 
shown in Fig.~\ref{ccxsec}b are
rather independent of the theoretical input and are in 
agreement with global fits.

Within the Standard Model, CC interactions are mediated by the
$t$-channel exchange of a $W$ boson, and therefore are sensitive to
the $W$ mass in the space-like regime.  
The propagator mass $M_{W}$ is fitted to the
double differential CC cross section data and yields
for $e^+p$ and $e^-p$ respectively\cite{zeushiq2,epub} 
$$M_{W}~=~81.4^{+2.7}_{-2.6} (\rm stat.)
            \pm 2.0 (\rm syst.)^{+3.3}_{-3.0} (\rm th.)~\rm GeV$$
\vspace{-1.0cm}
$$\hspace{0.5cm} M_{W}~=~79.9 \pm 2.2 (\rm stat.)
            \pm 0.9 (\rm syst.)  \pm 2.1 (\rm th.)~\rm GeV$$
These values are consistent with direct measurements in the
time-like domain from LEP and TEVATRON.  

\section{Conclusions}
The present tests of the Standard Model show good agreement 
of predictions of perturbative QCD and electroweak theory with the HERA data.
The luminosity upgrade of the HERA collider will bring
the opportunity to improve the precision of such tests, 
especially for the electroweak sector
at high $Q^2$ where the precision of the measurements 
is still limited by statistics.
%

%
\end{document}